\documentclass[twocolumn,fleqn]{svjour2}
\usepackage{graphicx,url}
\usepackage{amsmath,amsfonts,amssymb,amsopn,amscd}
\usepackage{mathptmx}
\usepackage[centerlast]{subfigure}
\usepackage{color}
\usepackage{colortbl}
\usepackage{epsfig}

\newcommand{\degC}{{\,}^\circ C}
\newcommand{\be}{\begin{equation}}
\newcommand{\ee}{\end{equation}}
\newcommand{\bea}{\begin{eqnarray}}
\newcommand{\eea}{\end{eqnarray}}
\def\avg#1{\left\langle{#1}\right\rangle}	% define brackets

\def\url#1{\textcolor{blue}{\underline{#1}}}	% URL in blue
\definecolor{violet}{rgb}{1.00,0.00,1.00}
\definecolor{turquoise}{rgb}{0.00,0.40,0.50}    % like "sea green"
\definecolor{lightred}{rgb}{0.9,0,0}		% light red

\journalname{Journal of Computational Neuroscience}

\emergencystretch=\maxdimen
\begin{document}

\renewcommand{\baselinestretch}{1.1} \small\normalsize

% ------------------------------------------------------------
% Title page
% ------------------------------------------------------------

\title{Evidence for frequency-dependent extracellular impedance from
the transfer function between extracellular and intracellular
potentials}

% Short title (if needed)
% \subtitle{Transfer function for LFPs}

\author{Claude B\'edard$^1$, Serafim Rodrigues$^1$, Noah Roy$^2$,
Diego Contreras$^2$ and Alain~Destexhe$^1$}

\institute{1: Integrative and Computational Neuroscience Unit (UNIC),
\\ UPR2191, CNRS, Gif-sur-Yvette, France \\
\email{Destexhe@unic.cnrs-gif.fr} \\ 2: Department of Neuroscience,
University of Pennsylvania, Philadelphia, USA.}

\date{\today}

\maketitle

% ------------------------------------------------------------
% Abstract
% ------------------------------------------------------------
% Please provide an abstract of 100 to 150 words. The abstract should
% not contain any undefined abbreviations or unspecified references. 

\begin{abstract}

We examine the properties of the transfer function $F_T = V_m /
V_{LFP}$ between the intracellular membrane potential ($V_m$) and
the local field potential ($V_{LFP}$) in cerebral cortex.  We first
show theoretically that, in the {subthreshold} regime, the
frequency dependence of the extracellular medium and that of the
membrane potential have a clear incidence on $F_T$.  The
calculation of $F_T$ from experiments and the matching with
theoretical expressions {is possible for desynchronized states
where individual current sources can be considered as independent. 
Using a mean-field approximation, we obtain a method} to estimate
the impedance of the extracellular medium without injecting
currents.  We examine the transfer function for bipolar
(differential) LFPs and compare to simultaneous recordings of $V_m$
and $V_{LFP}$ {during desynchronized states} in rat barrel cortex
{\it in vivo}.  The experimentally derived $F_T$ matches the one
derived theoretically, only if one assumes that the impedance of
the extracellular medium is frequency-depen\-dent, and varies as
$1/\sqrt{\omega}$ (Warburg im\-pedance) for frequencies between {3
and 500~Hz}.  This constitutes indirect evidence that the
extracellular medium is non-resis\-tive, which has many possible
consequences for modeling LFPs.

% removed:
% The power spectrum of $F_T$ can be non-monotonic and present a
% maximum which depends on the different time constants in the neuron
% membrane.  

\end{abstract}

% ------------------------------------------------------------
% Key Words
% ------------------------------------------------------------
% Please provide 4 to 6 keywords which can be used for indexing


% ------------------------------------------------------------
% Introduction
% ------------------------------------------------------------

\section{Introduction}

There is a widespread consensus that mechanisms for generating the
intracellular electrical activity are very well understood, however not
complete.  In contrast, {much less is known about the the genesis of
  extracellular potentials, which} is a subject of intense research. This is
associated to the difficulty in assigning measurements of the extracellular
potentials to a unique neurophysiological generator, which makes modeling of
LFP/EEG a complex issue.  Some of these mechanisms are believed to be related
to synaptic activity, synchronous population spikes, ephaptic interactions,
ionic dynamics, morphological structure of the neurons and many other
processes (reviewed in \cite{Jefferys1995,NunezSrinivasan2006}).

One of the characteristics of extracellular potentials is the very steep
attenuation of ``fast'' events such as spikes, which are visible only within
the immediate vicinity (a few microns) of the electrode.  In contrast,
``slow'' events such as synaptic potentials are visible for much larger
distances, typically a few hundred microns \cite{Des1999,Katzner2009}.  One
way to explain this differential filtering is that the extracellular medium
acts as a powerful low-pass filter \cite{Bed2004}.  However, this aspect is
controversial because some measurements of brain conductivity did not display
significant filtering effects \cite{Logo2007} while other measurements did
\cite{Ranck63,Gabriel1996}.  {Some of these experiments \cite{Gabriel1996}
  used careful controls, such as correcting for electrode polarization,
  showing the different frequency-dependence of various biological tissues,
  but most importantly, the independent measurements of conductivity and
  permittivity with theoretical constraints (Kramers-Kronig relations)
  \cite{Gabriel1996}.  The recent measurements employed a sophisticated
  four-electrode measurement setup to yield more accurate measures
  \cite{Logo2007}.  However, despite such controls,} these measurements used
current intensities that are much larger than biological sources, which may
explain the discrepancy \cite{BedDes2009a}.

In the present paper, we provide theoretical work and analyze experimental
measurements to examine whether the extracellular medium is non-resistive.  We
examine a quantity which depends on the extracellular impe\-dance, the
transfer function between simultaneously recorded intracellular and
extracellular potentials.  We show theoretically that, in the linear regime
and {for desynchronized states, the transfer function calculated from an
  (intracellular) single recording site} strongly depends on the extracellular
impedance, and can therefore be used to investigate its frequency dependence.
{We show preliminary results from desynchronized states {\it in vivo}}, which
indicate that the extracellular medium is indeed frequency dependent.  {We
  relate these findings to previous work and discuss their} possible
implications for modeling LFPs/EEG.

% ------------------------------------------------------------
%  Methods
% ------------------------------------------------------------

\section{Methods}

{The experimental data used in this paper were taken from a large database of
  cells \cite{Wilent2005a,Wilent2005b}, in which we selected intracellular
  recordings with long periods of subthreshold activity, simultaneous with LFP
  recordings in the vicinity (1~mm) of the intracellular electrode, and in
  light anesthesia with desynchronized EEG.

\subsection{Surgery and Preparation}

Experiments were conducted in accordance with the ethical
guidelines of the National Institutes of Health and with the
approval of the Institutional Animal Care and Use Committee of the
University of Pennsylvania.  Adult male Sprague-Dawley rats
(300-350g, n=35) were anesthetized with isoflurane (5\% for
induction, 2\% during surgery), paralyzed with gallamine
triethiodide, and artificially ventilated.  End tidal CO2
(3.5-3.7\%) and heart rate were continuously monitored.  Body
temperature was maintained at 37$\degC$ via servo-controlled
heating blanket and rectal thermometer (Harvard Apparatus,
Holliston, MA).  The rat was placed in a stereotaxic apparatus and
a craniotomy was made directly above the barrel cortex (1.0-3.0~mm
A/P, 4.0-7.0~mm M/L), and the dura was resected.  The cisterna
magna was drained to improve stability.  For intracellular
recordings, additional measures were taken to improve stability,
including dexamethasone (10~mg/kg, i.p.) to reduce brain swelling,
hip suspension, and filling the craniotomy with a solution of 4\%
agar.

\subsection{Electrophysiological Recordings}

Recordings of local field potentials (LFPs) across the cortical
depth were performed with 16-channel silicon probes (Neuronexus,
Ann Arbor, MI).  Probe recording sites were separated by {$100~\mu
m$} and had impedances of 1.5-2.0 M$\Omega$ at 1~kHz.  The probe
was lowered into the brain under visual guidance, oriented normal
to the cortical surface, until the most superficial recording site
was aligned with the surface.  {All neurons were regular-spiking
cells, and had spikes with about 2~ms width, so they were
presumably excitatory.  All LFP signals shown here were
{obtained by pairs of closely-located (400-500~$\mu$m apart)
electrodes arranged vertically} (surface-depth), and were amplified
and band-pass filtered at 0.1~Hz--10~kHz (FHC, Inc., Bowdoinham,
ME).}

Intracellular recordings were performed in barrel cortex with glass
micropipettes pulled on a P-97 Brown Flaming puller (Sutter Instrument
Company, Novato, CA).  Pipettes were filled with 3M potassium acetate and had
DC resistances of 60-80~M$\Omega$.  A high-impedance amplifier (low-pass
filter of 5~kHz) with active bridge circuitry (Neurodata, Cygnus Technology,
Inc., Delaware Water Gap, PA) was used to record and inject current into
cells.  Vertical depth was measured by the scale on the micromanipulator.  A
Power 1401 data acquisition interface and Spike2 software (Cambridge
Electronic Design, Cambridge, U.K.) were used for data acquisition.

\subsection{Analysis and simulations}

All simulations and analyses were realized using MATLAB (Mathworks
Inc, Natick, MA), {except for compartmental model simulations
(see below).}

{The analysis of the PSD was performed using two different methods.  A
  constrained nonlinear least square fit of the analytic expressions for
  different possible transfer functions (see Eq.~\ref{WarburgDiffFT} and
  Eq.~\ref{RC_resisitveFT} in Results) was performed to the transfer function
  calculated from experimental data.  The fit was constrained to frequencies
  between 3~Hz and 500~Hz and the parameters of the transfer functions were
  constrained to the physiological range (see Results for details).}

{Because the PSD and calculated transfer functions had large
variance, two different methods were used to perform the fitting. 
First, a polynomial averaging algorithm was used to calculate the
mean value of the PSD, and the fit was performed against this mean
value (see details in Appendix~B).  Second, a moving average window
procedure which consisted of partitioning the data set into five
epochs (corresponding to about 7.7~sec for one cell and about
4.4~sec for the second cell, and additionally applied a Hanning
window to each epoch (a Hamming window was tested as well, and gave
similar results).  This method markedly reduces the variance of the
PSD and of the transfer function.  Other choices for window length
were also tested and yielded similar results as those shown in the
figures (not shown).}

{\subsection{Compartmental model simulations}}

{Compartmental models were simulated using the
freely-available NEURON simulation environment~\cite{Hines1997}.  A
16-compartment ``ball-and-stick'' model was simulated, based on a
soma (area of 500~$\mu$m$^2$) and 15 dendritic compartments of
equal length (46.6~$\mu$m) and tapering diameter (from 4~$\mu$m at
the soma up to 1~$\mu$m at the distal tip), so that the total cell
area was of 6000~$\mu$m$^2$.  Each compartment had passive
properties, with axial resistivity of 250~$\Omega \cdot $cm,
specific capacitance of 1~$\mu$F/cm$^2$, resting membrane
conductance of 0.45~mS/cm$^2$ and reversal potential of -80~mV.}

{To simulate high-conductance states similar to {\it in vivo}
measurements, each dendritic compartment contained two randomly
fluctuating synaptic conductances (model from~\cite{Des2001}) which
were adjusted so that the mean excitatory and inhibitory
conductances were respectively of 0.73 and 3.67 times the resting
conductance.  These values represent mean values of the
``spontaneous'' synaptic activity during activated states {\it in
vivo}, according to intracellular measurements (see details in
\cite{Des2003}).  We verified that using these parameters, the
subthreshold V$_m$ activity of the cell was conform to typical
intracellular measurements during desynchronized-EEG states, as for
example in the experiments reported here.}

{In a separate set of simulations, the synaptic inputs were 
present only in the most distal compartment.  In this case, all 
synaptic conductances were set to zero in all other compartments.}

{The LFP generated by this compartmental model was calculated at
{the level of the stem dendrite, 50~$\mu$m lateral to the dendritic
axis}, assuming a purely resistive medium, according to the
expression:
\begin{equation}
  V_{LFP} \ = \ {R_e \over 4 \pi} \ \sum_{j} { I_j \over r_j },
\end{equation}
where $R_e$ = 230~$\Omega \cdot$cm is the extracellular
resistivity~\cite{Ranck63}, $I_j$ is the total membrane current of
compartment $j$, and $r_j$ is the distance between compartment $j$
and the extracellular position where the LFP is evaluated.}
{The LFP of the full model was compared to the LFPs calculated
from each individual current source similar to above, with
$V_{LFP}^{(j)} \ = \ {R_e \over 4 \pi} \ { I_j \over r_j }$.} \\

% ------------------------------------------------------------
%  Theory
% ------------------------------------------------------------

\section{Theory}

In this section we formulate a theoretical relationship between extracellular
and intracellular sub-threshold voltage activity. {We consider a linear
  electromagnetic regime, which is defined by the fact that the linking
  equations are linear in frequency space.  In such case we have
  {$\vec{j}_{f} = \sigma_{f}\vec{E}_{f}$, $\vec{D}_{f} =
    \epsilon_{f}\vec{E}_{f}$ and $\vec{B}_{f} = \mu_{f}\vec{H}_{f}$ where the
    second-order symmetric tensors $\sigma_{f}$, $\epsilon_{f}$ and $\mu_{f}$
    do not depend on $\vec{E}_{f}$ and $\vec{H}_{f}$. $\sigma_{f}$,
    $\epsilon_{f}$ and $\mu_{f}$ depend on position in general.}  In other
  words, we will work in the subthreshold voltage range, for which we assume
  linear current-voltage relations, and that conductances are not voltage
  dependent.  Henceforth, we will refer to this dynamical state as ''linear
  regime''.} {We also assume that the cortical tissue is macroscopically
  isotropic, in which case the tensors become scalar functions.}

We will also formulate the model in the frequency domain, to easily
enable comparison between the amount of intracellular and
extracellular signal within each frequency band, which is readily
accessible from experimental data.  The advantage of this approach
is simplicity as it avoids tackling directly the different time
scales inherently associated to LFP/EEG signals, many of which are
not yet understood.  For example, it is known that LFP/EEG signals
are composed of mixed-mode oscillations, however the origins and
transitions between these modes are not yet explained
\cite{Erchova2008}. The choice of working in the frequency domain
also seems natural because the differential equations describing
electric potentials transform into algebraic equations in frequency
space.

{The formulation of the relation between transmembrane ($V_m$) and
extracellular ($V_{LFP}$) activity requires to derive the transfer
function, $F_T$, that measures the ratio between the impedance of
the cellular membrane and extracellular medium. To obtain this, we
start from a general formalism where $V_{LFP}$ is expressed as a
function of a large number of current sources located at different
positions in extracellular space (Section~\ref{gen}).  We then show
under what conditions this formalism can reduce to a single current
source, and derive the impedance for this current source
(Section~\ref{imped}) and derive the corresponding transfer
functions (Sections~\ref{ft} to \ref{diff}).}

In the following, we use the notation {$F(f)$ or $F_f$} to denote the
function $F$ in Fourier space {with frequency $f$ and $\omega = 2 \pi
  f$.}

% ------------------------------------------------------------
%  subsection
% ------------------------------------------------------------
{\subsection{General model with multiple current sources}} \label{gen}

{We start from the general model in Fourier space:
{
\begin{equation}
 V_{LFP}(r_i,f) = \sum_{j=1}^{N} \frac{R_{j}}{d_{ij}} Z_j^{med}(f) I_j(f)  ~ ,  
 \label{geneq}
\end{equation}}
{where $V_{LFP}(r_i,f)$ is the extracellular potential at a
position $r_i$ as resulting from a set of $N$ monopolar current
sources $I_j(f)$.  $R_{j}$ is a constant, $d_{ij}$ is the distance
between source $j$ and position $r_i$, and $Z_j^{med}$ is the
impedance of the extracellular medium around source $j$.}  This
model is based on the property that any charge distribution in
space and frequency can be expressed as a sum of monopolar sources.
This therefore applies to complex current distributions in the
complex dendritic structures of many neurons surrounding the
recording site $r_i$ (see Fig.~\ref{schem}A), and thus, this
formalism is general.}

%-------------------------------------------------------------------------------------------------------------------------------------
%    Figure
%-------------------------------------------------------------------------------------------------------------------------------------
\begin{figure}[h!]
\centering
\includegraphics[width=0.9\columnwidth]{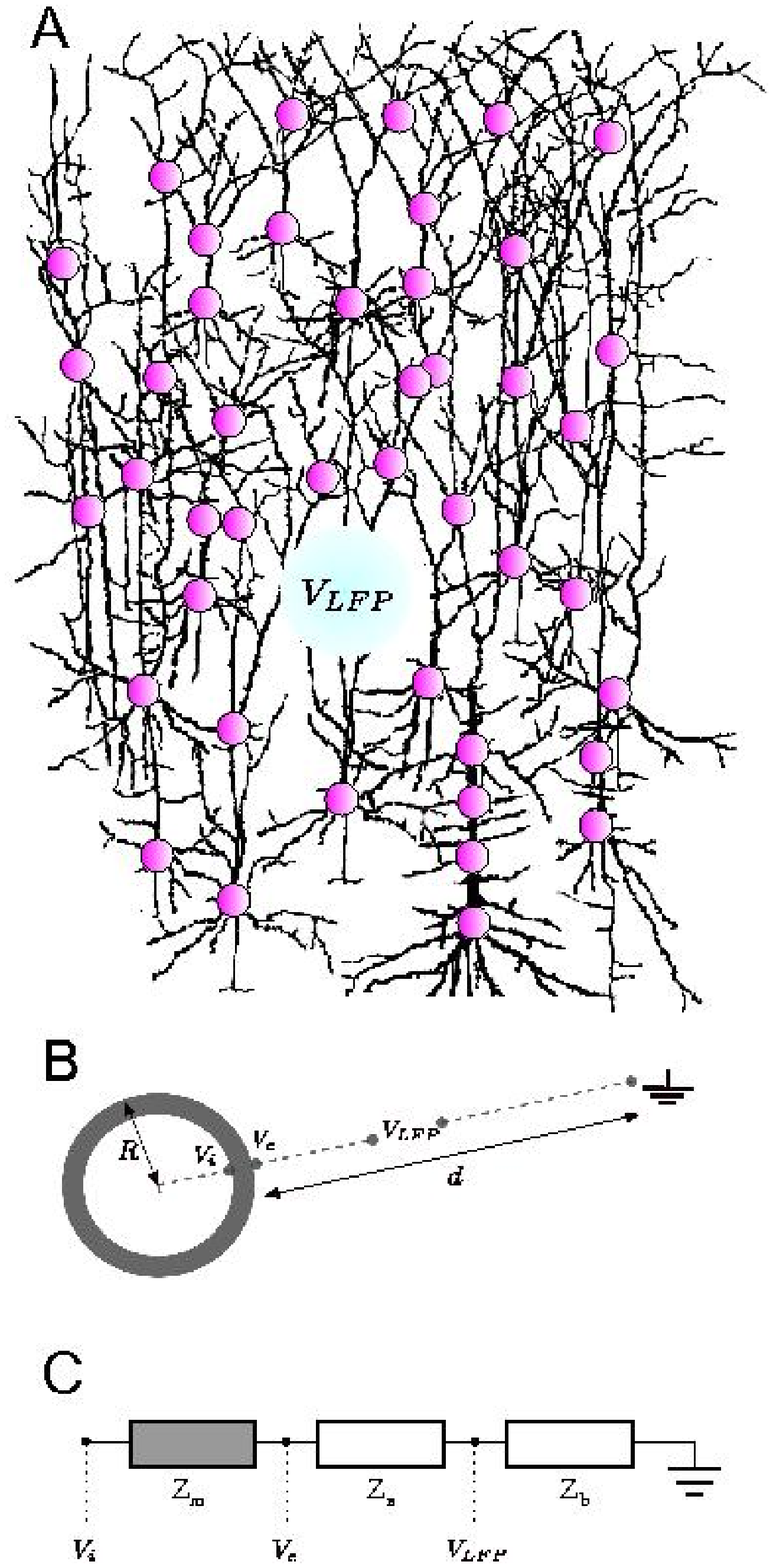}

\caption{Arrangement of potentials and impedances. {A. Scheme of
the general model where the local field potential ($V_{LFP}$) is
generated by a large number of individual monopolar current sources
(red) distributed in soma and dendrites of neurons in the network. 
If the activity of these current sources is desynchronized, the
system can reduce to a single current source using a mean-field
approximation.}  {B. Scheme of an individual current source, which
is assumed to be of spherical symmetry with homogeneous
extracellular medium, so that there is no position-dependence of
conductivity or permittivity. $V^i$ and $V^e$ are the electric
potentials, respectively inside and outside of the membrane,
relative to a reference potential $R_f^{\infty}=0$ situated at an
infinite distance.  The membrane potential $V_m$ is the difference
between $V_i$ and $V_e$.} $V_{LFP}$ is the local field potential,
which is the voltage difference between a point $P$ in
extracellular space and $R_f^{\infty}$.  C. Equivalent electrical
circuit for each current source, where $Z_m$ is the impedance of
the membrane, while $Z_a$ and $Z_b$ are impedances of the
extracellular medium.}

\label{schem}
\end{figure}
%-------------------------------------------------------------------------------------------------------------------------------------

{If we assume that the extracellular medium is electrically homogeneous, the
  extracellular impedance $Z_j^{med}$ is the same for every pair of points
  $i,j$, and the local field potential becomes: {
\begin{equation}
 V_{LFP}(r_i,f) = Z^{med}(f) \sum_{j=1}^{N} \frac{R_{j}}{d_{ij}} I_j(f)
 \label{geneq2}
\end{equation}}}

{We also assume that the network activity is of low correlation, as
typically found during desynchronized network states {\it in vivo},
such as in awake animals, or during the ``up-states'' of
anesthesia.  In such states, it was shown that the activity is very
irregular with low levels of correlation between cortical neurons
\cite{Des1999,Gawne93,Steriade2001b,Zohary94}.  As a consequence,
we can consider that the dendritic structure is bombarded by noisy
synaptic events that are essentially uncorrelated, and if we assume
that this activity primes over the deterministic link between
individual current sources {(see {Section~\ref{test}} for a
test of this assumption)}, then the field produced by the ensemble
of $N$ current sources is equivalent to the field produced by
independent sources. the variables $\frac{R_{j}}{d_{ij}}$ and $I_j$
can be considered as statistically independent, and by averaging
over the ensemble of current sources, we obtain: 
{\begin{equation}
 V_{LFP}(r_i,f) = Z^{med}(f) \sum_{j=1}^{N} \frac{R_{j}}{d_{ij}} 
 I_j(f) = Z^{med}(f) \ \bar{R}_D \ \bar{I}(f) ~ .  
\end{equation}}
Here, the LFP has been expressed as a function of a ``mean current
source'' (in the spatial sense) {$\bar{I}(f) = N <I_j(f)>|_j$} and
where {$\bar{R}_D$} = $<\frac{R_{j}}{d_{ij}}>|_j$
\footnote{{Note that for desynchronized activity, we expect that
such a spatial average will be of small amplitude, as indeed
typically found for the ``desynchronized EEG'' condition
investigated here.}}.  This formulation is equivalent to a
mean-field theory expressed in Fourier space and where we consider
the LFP as generated by the mean current source contribution.}

% ------------------------------------------------------------
%  subsection
% ------------------------------------------------------------
\subsection{{Testing the assumptions of the formalism}}
\label{test}

{To verify some of the assumptions of the formalism, we have
simulated a ball-and-stick dendritic model subject to fluctuating
synaptic conductances (Fig.~\ref{desynch}).  In a first set of
simulations, the synaptic conductances (model from \cite{Des2001})
were located only at a single dendritic location.  This case
produced LFPs which were markedly filtered by the dendritic
structure (Fig.~\ref{desynch}A, middle panel), as described
previously \cite{PetEin2008}.  This type of filtering is due to the
fact that in dendritic structures, the return current (which
participates to generate the LFP) is filtered by the membrane.  In
another set of simulations, we have considered a more realistic
situation in which the synaptic conductances were located at all
locations of the dendritic structure, so that the total conductance
matches that measured experimentally {\it in vivo} \cite{Des2003}. 
In this case, the filtering due to morphology was negligible
(Fig.~\ref{desynch}B, middle panel), presumably because synaptic
currents largely dominate over axial currents.  These simulations
suggest that the filtering due to morphology can be neglected for
{\it in vivo}--like conditions, {and that in such conditions,
the filtering must come from another origin, such as the
extracellular medium}. }

%-------------------------------------------------------------------------------------------------------------------------------------
%    Figure
%-------------------------------------------------------------------------------------------------------------------------------------

\begin{figure}[h!]
  \centering
\includegraphics[width=\columnwidth]{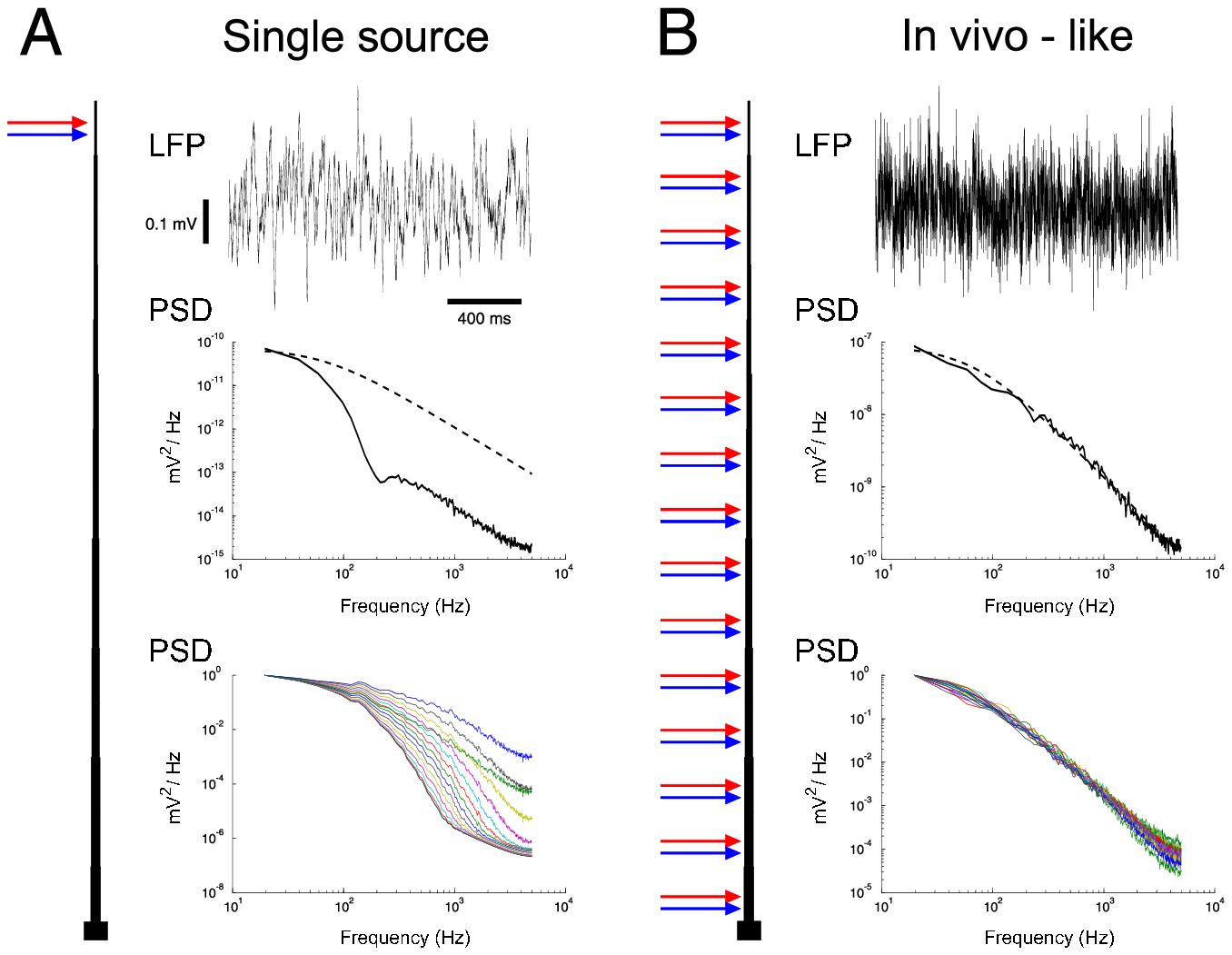}

\caption{{Test of the assumptions of the formalism using a
ball-and-stick model.  A. Top: scheme of the model.  Excitatory
(blue) and inhibitory (red) random synaptic conductances (model
taken from \cite{Des2001}) were inserted in the distal compartment
of a ball-and-stick model with 16 compartments (see Methods). 
Middle: LFP calculated from the total membrane currents, at
50~$\mu$m from the neuron.  {Middle}: PSD of the LFP, which
showed a marked filtering due to the neuronal morphology, as
described previously \cite{PetEin2008}.  {Bottom: PSDs
obtained from the LFP of each individual source.} B. Same
simulation as in A, but the random synaptic inputs were distributed
in all dendritic compartments and were matched to the the
conductance state observed {\it in vivo}.  {Middle}: the
filtering due to morphology was negligible in this case. 
{Bottom: individual PSDs are very similar, suggesting that a
mean-field approximation is justified in this case.} The dotted
curves represent the optimal fit to the PSD in B, using a
Lorentzian function $1 / (1 + \omega^n \tau^n)$ with $n=1.8$ and
rescaled to the maximum of each PSD.}}

\label{desynch} 
\end{figure}

%-------------------------------------------------------------------------------------------------------------------------------------

{To test the validity of a mean-field approximation, we have
plotted the PSD of the LFP generated by each individual source in
the ball-and-stick neuron (Fig.~\ref{desynch}A,B, bottom panels). 
In the case the synaptic current is located at a single dendritic
location, the PSDs of individual sources are all different
(Fig.~\ref{desynch}A, bottom), because of the filtering due to the
morphology.  However, in the case of distributed current sources,
the PSDs of different current sources are very similar
(Fig.~\ref{desynch}B, bottom).  This similarity also extended to
LFPs calculated from different locations in extracellular space,
which produced LFPs of different amplitudes, but with almost
superimposable PSDs (not shown).  These simulations suggest that,
for in {\it in vivo}--like states, a mean-field approximation is
reasonable.}

% ------------------------------------------------------------
%  subsection
% ------------------------------------------------------------
\subsection{Membrane impedance for a spherical source in the linear
regime} \label{imped}

{In the previous sections, we {assumed} that the LFP generated in
desynchronized network states has equivalent properties to that
produced by a ``mean current source''.  We now calculate the
membrane impedance for an individual current source of spherical
symmetry,} embedded in a medium which is homogeneous/isotropic and
continuous, and with electric parameters {$\sigma_f$},
{$\epsilon_f$}\footnote{Note that we keep the electric parameters
frequency dependent, to keep the expressions as general as
possible.  In addition, the theory can easily be generalized to
multipoles, as any multipole configuration can be decomposed in a
sum of monopoles, and to multiple sources using the linear
superposition principle.}.  The cell is represented by an RC
circuit where the membrane potential is expressed as the difference
{$V_m = V_i-V_e \approx -70~ mV$} (Fig.~\ref{schem}B).  {At the
resting membrane potential}, there is a net negative charge $Q_-$
inside, which is perfectly balanced with a net positive charge
$Q_+$ on the external surface of the membrane, {such that} $Q_- +
Q_+ =0$.  In this situation, the electric field $\vec{E}$ produced
by the cell in the medium is null, which implies that {$V_e =
V_f^{\infty} = 0$, where $V_f^{\infty}$} is the reference potential
at infinite distance from the source.

Suppose that a small excess of positive charge is injected inside the cell.
Due to this excess of charge, an electric field $\vec{E}$ will instantaneously
{appear} in extracellular space.  {An electric current will also appear}
to restore the equilibrium (and therefore {will give rise to} a variation of
the electric field). The current produced depends on the physical and
biological characteristics of the membrane, which will determine the time
evolution of the electric field.  The current is given by:
\begin{eqnarray}
 I_r &=& \sum_{i=1}^{N} g_i~(t,V_m) (V_m(t)-E_i)\nonumber  \\ \nonumber \\
 I_c &=& C_{m}\frac{d V_m}{dt}\nonumber \\ \\
 I_m &=& I_r +I_c \nonumber \\\nonumber  
\end{eqnarray}
where $I_r$ and $I_c$ are the ionic and capacitive currents, respectively.
The index $i$ represents the different membrane conductances $g_i$, with their
reversal potential $E_i$, and $C_{m}$ is the membrane capacitance.

In the following, we will assume that the excess of charge remains small and
varies around a stationary mean value, so that we can can neglect the time
variations of conductances and of the $V_m$ ($\Delta V_m$), with ${\partial
  g_j^i}/{\partial V_m}\approx 0$.  In this case, we can consider that the
conductances are only function of the mean value of the $V_m$.  This is
equivalent to assume that, in the subthreshold regime, the conductances are
not dependent on the potential, and that the reversal potentials $E_i$ are
constant.  This is valid if the impact of ionic concentration changes on $E_i$
are negligible compared to the voltage variations.

Under these approximations, we can write:
\begin{eqnarray}
 \Delta I_r &=&
 \sum_{i=1}^{N}g_i~(\avg{V_m}|_t)
  \Delta V_m\nonumber  \\ \nonumber \\
 \Delta I_c &=& C_{m}\frac{d \Delta V_m}{dt}\nonumber \\ \\
 \Delta I_m &=& \Delta I_r +\Delta I_c \nonumber \\\nonumber  \\
\end{eqnarray}
Here, we have a linear system of equations with time-independent coefficients.
By expressing the variation of current produced by the cell as a function of 
the variation of membrane voltage, in Fourier space, we obtain:
{
\begin{eqnarray}
 \Delta I_r(f) &=& G_m  \Delta V_m(f)\nonumber  \\ \nonumber \\
 \Delta I_c(f) &=& i \omega C_{m} \Delta V_m(f)\nonumber \\ \\
 \Delta I_m(f) &=& \Delta I_r(f) +\Delta I_c(f) \nonumber \\\nonumber  
\end{eqnarray}}
where 
$$G_m=\sum_{i=1}^{N}g_{i} ~ . $$

The impedance of the membrane is given by:
{
\begin{equation}
 Z_m(f)=\frac{\Delta V_m(f)}{\Delta I(f)} = 
 =\frac{R_m}{1+i\omega \tau_m}
\label{Zm}
\end{equation}
where $R_m =\frac{1}{G_m}$.}

{We next define the theoretical transfer function, which provides a
relation (in this case, in the frequency domain) between LFP and
{$V_m$}, as {$F_T(\vec{r},f)=\frac{V_m}{V_{lfp}}(\vec{r},f)$}.
Having defined above the impedance of the membrane we now require
to define the impedance of the medium to fully characterize the
transfer function. Consequently, in subsequent sections, we define
the transfer function in general form and we then consider
particular cases due to different recordings montages (bipolar and
monopolar) and various types of media.}

% ------------------------------------------------------------
%  subsection
% ------------------------------------------------------------
\subsection{{Computing the transfer function of each current source.}} 
\label{ft}

{In this section, we calculate separately the two transfer functions:
{
$$
F_T^{(1)}(f)=\frac{V_m(f)}{V_e(f)}
$$ and $$
F_T^{(2)}(d,f)=\frac{V_e(f)}{V_{LFP}(d,f)} ~ .
$$}}

{{$F_T^{(1)}(f)$\footnote{{Note that the notation $F_T^{(n)}$ stands
        for different functions and not for $n^{th}$ order derivative.}}} can
  be calculated from the equivalent circuit shown in Fig.~\ref{schem}C.  We
  have the following relations: {
\begin{eqnarray}
Z_{med}(R,f) &=& Z_a(d,f) + Z_b(d,f) \\
V_m(f) &=& V_i(f) - V_e(f) \\
\frac{V_i(f)}{Z_m(f) + Z_{med}(R,f)}   &=& \frac{V_e(f)}{Z_{med}(R,f)} 
\end{eqnarray}}
where $R$ is the radius of the current source and $d$ is the distance
relative to the center of the source.  It follows that
{
\begin{equation}
F_T^{(1)}(f)=\frac{V_m(f)}{V_e(f)}
=\frac{Z_m(f)}{Z_{med}(R,f)}
\end{equation}}}

{{$F_T^{(2)}(d,f)$} can also be calculated based on the equivalent 
circuit of Fig.~\ref{schem}C:
{
\begin{equation}
F_T^{(2)}(d,f)=\frac{V_e(f)}{V_{LFP}(d,f)}
=\frac{Z_a(f)+Z_b(d,f)}{Z_b(d,f)}
\end{equation}}}

We note that if $Z_a$ and $Z_b$ have the same frequency dependence, for
example $f^n$, then $F_T^{(2)}$ is independent of frequency when the medium is
macroscopically homogeneous.  For example, if both media have a Warburg
impedance ({$Z \sim 1/\sqrt{f})$} or capacitive impedance ({$Z \sim
  1/f$}), the function $F_T^{(2)}$ has the same value as for the case of a
resistive medium (frequency independent impedance).  {For a spherical source
  and for an extracellular position at a distance $d$ from the center of the
  source, we have: {
\begin{equation}
F_T^{(2)}(\vec{r},f)=
\frac{V_e(f)}{V_{LFP}(\vec{r},f)}=\frac{d}{R}
\end{equation}}
where $R$ is the radius of the current source.  In this relation,} 
the resistance of a spherical shell of radius $r$ {in an isotropic and
  homogeneous medium of infinite dimension} equals $\frac{1}{4\pi\sigma r}$, 
which corresponds to the sum $Z_a +Z_b$ when $r=R$ and $Z_b$ when $r=d$.

In the case of a heterogeneous isotropic medium, we
have \cite{BedDes2009a}:
{
\begin{equation}
Z(r,f)=\frac{1}{4\pi\sigma_z(R)}
\int_{r}^{\infty}dr'\frac{1}{r'^2}~
\frac{\sigma_{f}(R)+i\omega~\epsilon_{f}(R)}
{\sigma_{f}(r') + i\omega~\epsilon_{f}(r')}
\label{impedance}
\end{equation}}
for a spherical and isopotential source. $\sigma_z$ represents the
complex conductivity. In the absence of spherical symmetry and with
non-isopotential sources, it is {necessary}, in general, to solve
differential {or integral} equations derived from Maxwell equations in the
quasi-static regime (neglecting electromagnetic induction; see
\cite{Chari1999}). 

{It is important to note that the expressions above for the transfer
  function are independent of the particular frequency spectrum of the current
  sources.}

% ------------------------------------------------------------
%  subsection
% ------------------------------------------------------------
\subsection{Frequency dependence of differential (bipolar) recordings}
\label{diff}

{If the power spectral density (PSD) of the LFP signal varies as}
$1/f^{\gamma}$ where $\gamma\geq 1$ (for monopolar LFP recordings),
then the energy associated to the signal would necessarily be
infinite, which is of course physically impossible.  In fact,
{according to a model developed previously (see Eqs.~53--54 in
\cite{BedDes2009a}), a more accurate relationship between the
extracellular medium's properties and the mean $V_f$ is given by
the following expression
\begin{equation}
 V(\vec{r},f)=
 \frac{\kappa_i(\vec{r})}{f^{\gamma/2}+ a_i} ~ .
\end{equation}
where $a_i$ is negligible for frequencies larger than 1~Hz and smaller
than about 500~Hz (because LFPs are usually considered up to 500~Hz).}  Note
that the PSD is proportional to the square of {$V({f})$} and will thus
scale as $1/f^{\gamma}$ in this case.  {The PSD is also independent of the
  position (homogeneous medium)}.

The values of constants $\kappa_i$ and $a_i$ represent respectively the
proportionality constant for each electrode $i$, which depends on the
intensity of the field for large frequencies, and the natural limit of the
value of the voltage for very low frequencies (which limits the energy of the
system).  In general, these constants depend on electrode position, and
therefore when one takes the difference between two electrodes, we have:
{
\begin{equation}
  V_{diff}(f)=V_{LFP}^{(1)}(f)-V_{LFP}^{(2)}(f)
  = \frac{\kappa_1}{f^{\gamma/2}+a_1}-\frac{\kappa_2}{f^{\gamma/2}+a_2}.
\end{equation}}

If the signal intensities of the two electrodes are comparable for
large frequencies, we have necessarily $\kappa_1 \approx \kappa_2$,
{such that the differential or bipolar signal (difference between
two nearby extracellular electrodes {(very correlated
signals)}, will have the following form, on average: 
{
\begin{equation}
  V_{diff}(f)\approx
    \kappa_1\cdot\frac{a_2-a_1}{(f^{\gamma/2}+a_2)(f^{\gamma/2}+a_1)}
    \approx  \frac{\hat{\kappa}}{f^{\gamma}}.
\label{diffeq}
\end{equation}}
where $\hat{\kappa} = \kappa_1(a_2-a_1)$ and for {$f>1Hz$ and
  $f<500~Hz$}.  Thus, if monopolar LFPs have a PSD which varies as
$1/f^{\gamma}$, one can have a PSD in $1/f^{2\gamma}$ in bipolar recordings.

% ------------------------------------------------------------
%  Numerical Simulations
% ------------------------------------------------------------
\section{Numerical simulations of transfer functions}

In this section, we present simple numerical simulations to illustrate how the
transfer function is influenced by the frequency dependence of the cellular
membrane, that of the medium, and of the recording configuration.

% ------------------------------------------------------------
% subsection
% ------------------------------------------------------------
\subsection{Resistive membrane with homogeneous/isotropic resistive
medium}

As a first and simplest case, suppose we have a resistive membrane (the
membrane capacitance is neglected), embedded in a homogeneous resistive
medium.  In this case, the resistive medium is described by Laplace equation,
and we have:
{
\begin{equation}
 F_T(f) = F_T^{(1)}(f)\cdot F_T^{(2)}(f)=\frac{R_m}{R_{med}}\cdot\frac{R}{d},
\end{equation}}
where $R$ is the radius of the source, $R_m$ is the membrane
resistance, $R_{med}$ is the resistance of the medium, and $d$ is
the distance from the LFP measurement site to the center of the
source.

This case, however, is not very realistic because the membrane capacitance is
neglected.  In the following sections, we consider more elaborate membranes
and different extracellular media.

% ------------------------------------------------------------
%  subsection
% ------------------------------------------------------------
\subsection{Capacitive effects of membranes in homogeneous/isotropic
resistive media}

We now consider a membrane with capacitive effects described by a simple RC
circuit, together with a resistive medium.  In this case, we have: 
{
\begin{equation}
 F_T(f) = F_T^{(1)}(f)\cdot F_T^{(2)}(f)=\frac{Z_m(f)}{R_{med}}\cdot\frac{R}{d},
\end{equation}}
where the parameters are as described above, with {$Z_m(f)$} the membrane 
impedance.  Thus, according to Eq.~\ref{Zm}, we have the following
transfer function:
{
\begin{equation}
 F_T^s(f) = \frac{R_m}{R_{med}}\cdot\frac{1}{1+i\omega \tau_m}\cdot\frac{R}{d}.
 \label{RC_resisitveFT}
\end{equation}}
{where $\tau_m$ is the membrane time constant.}  This transfer
function is depicted in Fig.~\ref{Models}A.  

The transfer function can also be calculated for a ``non ideal'' membrane,
with a more realistic RC circuit model where the capacitance is non-ideal and
does not charge instantaneously (see details in \cite{BedDes2008}; {see also
  Appendix~A}).  Considering such a non-ideal membrane with a resistive
medium, we have: {
\begin{equation}
 F_T^{N}(f) = \frac{R_m}{R_{med}}\cdot\frac{1}{1+i\frac{\omega \tau_m}{1+i\omega \tau_{\mbox{\begin{tiny}MW\end{tiny}}}}}\cdot\frac{R}{d}
\label{nst}
\end{equation}}
{where $\tau_{\mbox{\begin{tiny}MW\end{tiny}}}$ is the Maxwell-Wagner
  time of a non-ideal capacitance.  Note that when
  $\tau_{\mbox{\begin{tiny}MW\end{tiny}}}=0$}, we recover the case above for
an ideal membrane. This transfer function is represented in
Fig.~\ref{Models}A. The transfer function is in general monotonic (and scales
close to $1/f^2$).  

Interestingly, there is a phase resonance for non-ideal membranes (see
($\star$) in Fig.~\ref{Models}B). {The physical origin of this phase
  resonance in the non-ideal cable is that the membrane is quasi-resistive at
  low ($\simeq 0$) and high frequencies (100~Hz) (see Eq.~\ref{ZmN}), so that
  the absolute value of the phase must necessarily pass through a
  maximum\footnote{{Every positive continuous function defined on a
      compact domain has necessarily a maximum inside that domain.}}}

% ------------------------------------------------------------
%  subsection
% ------------------------------------------------------------

\subsection{RC membrane in homogeneous/isotropic non-resistive
medium\label{aa}} 

We now focus on non-resistive media by providing a transfer function for which
the functional form should be observable from extracellular measurements. We
consider the simplest case scenario of only linear subthreshold regime where
the membrane is described by a simple RC circuit embedded in a medium with
impedance {$Z(f)$}.  In this case, the transfer function for a mono-polar
extracellular recording is given by: {
\begin{equation}
 F_T(f) =  R_m \cdot
 \frac{1}{ Z(f) \ (1+i\omega \tau_m)}\cdot\frac{R}{d},
 \label{GeneralMonoFT}
\end{equation}}

{{If the impedance of the medium is $\frac{\kappa}{f^{\gamma/2}-a_i}$ and
    $\gamma \geq 1$, where $\kappa$ is a complex constant, the monopolar and
    bipolar transfer functions are respectively:}}
%-------------------------------------------------------------------------------------------------------------
{{
\begin{large}
\begin{equation}
\centering
\begin{array}{ccc}
 F_T^{mono}(f) &\approx & \frac{R_m}{\kappa}\cdot\frac{(f)^{\gamma/2}}{1+i\omega \tau_m}\cdot\frac{R}{d}.\vspace{2mm}\\
F_T^{diff}(f) &\approx & \frac{R_m}{\hat{\kappa}}\cdot\frac{f^{\gamma}}{1+i\omega \tau_m}\cdot\frac{R}{d}.
 \label{WarburgMonopolarFT}
\end{array}
\end{equation}
\end{large}}}
%--------------------------------------------------------------------------------------------------------------

{The differences between monopolar and bipolar transfer functions are
  explained in Section~\ref{diff}.} In Fig.~\ref{Models}C, we show few
examples of such a case with different values of the membrane time constant
$\tau_m$.  We observe that the modulus of the transfer function can present a
maximum which depends on $\tau_m$, and therefore of the level of activity or
the ``conductance state'' of the membrane.

%-------------------------------------------------------------------------------------------------------------------------------------
%    Figure
%-------------------------------------------------------------------------------------------------------------------------------------
\begin{figure}[h!] \centering
\includegraphics[width=0.8\columnwidth]{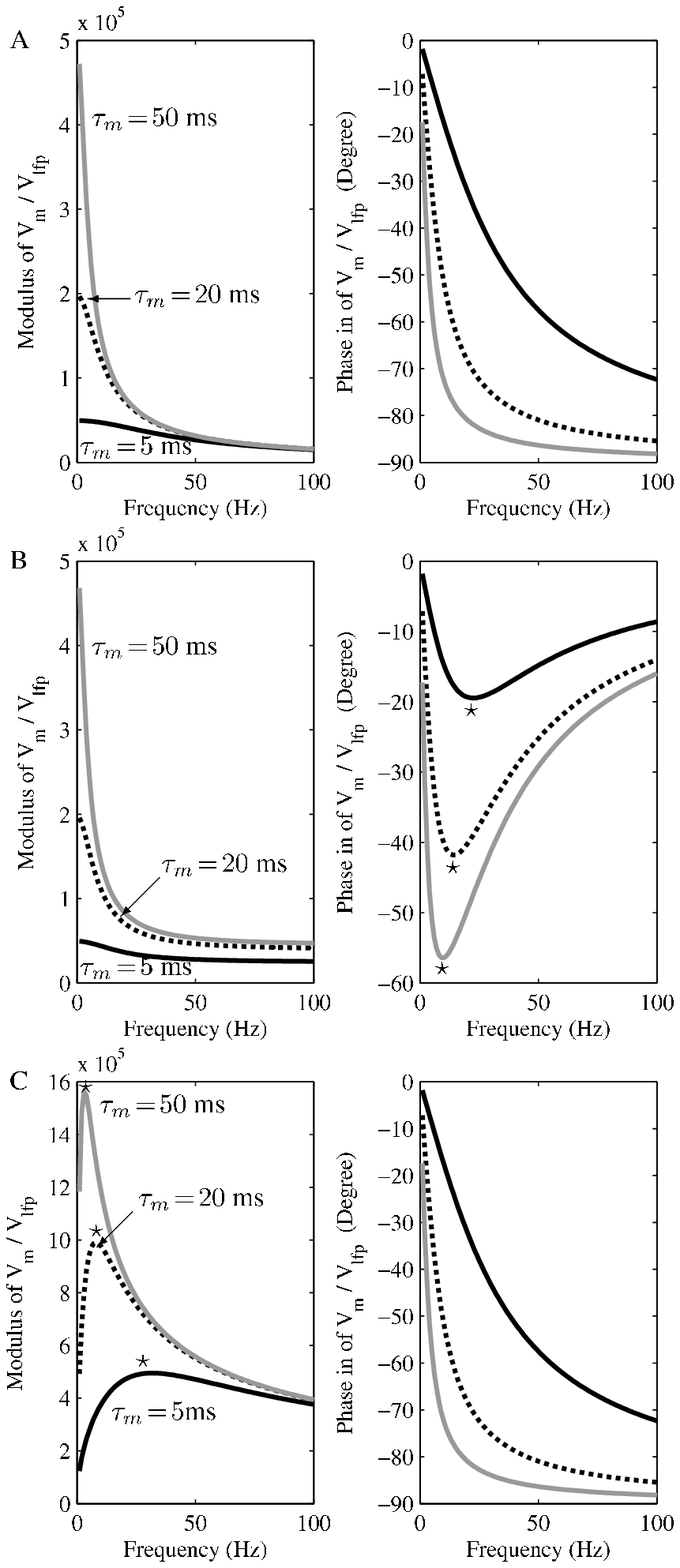}

\caption{Amplitude and phase of the transfer function $F_T$ as a function of
  frequency for different models and for mono-polar electrode montage. In all
  cases the transfer function was estimated for a distance of $30~\mu m$ from
  a spherical source of 10~$\mu m$ radius.  A. Standard RC membrane model with
  $C_m=10^{-2}~F/m^2$ and various configurations for membrane time constant
  $\tau_m$. The extracellular medium conductivity is $0.3~S/m$.  B.  Non-ideal
  membrane model (see \cite{BedDes2008}) with same parameters as in A, and
  with a Maxwell-Wagner time $\tau_{MW}=5~ms$ {(see Appendix~A)}. The
  extracellular medium is also resistive in this case.  Surprisingly, we
  observe a resonance in the phase of the transfer function as indicated by
  ($\star$).  C. Standard RC membrane model together with a medium described
  by a Warburg Impedance {$\frac{e^{i\phi}}{4\pi\sigma\sqrt{f}}$ where $\phi$
    is frequency independent. We have taken $\phi =0$ to illustrate the
    differences between this type of impedance compared to a resistive medium.
    $\phi$ will be different than zero in general, with no change to the
    modulus.} Note that there is a peak in amplitude of the absolute value of
  the transfer function that increases and shifts to lower frequencies as
  $\tau_m$ increases; $\star$ indicates a resonance. }

\label{Models}
\end{figure}
%-------------------------------------------------------------------------------------------------------------------------------------

% ------------------------------------------------------------
%  Comparison with experimental results
% ------------------------------------------------------------
\section{Comparison with experimental results} \label{exp}

In this section, we test the resistive or non-resistive nature of the
extracellular medium by evaluating the transfer function from experimental
data and compare with theoretical estimates. However, we note that in our
experiments we use bipolar LFPs recorded in rat barrel cortex, simultaneously
with intracellular recordings in the same cortical area.  Since the LFP
recordings are bipolar, {we use the following form (see Section~\ref{aa}):}
{
\begin{equation}
 F_T^{(diff)}(f) = 
\frac{R_m ~f^{\gamma}}{\hat{\kappa}\ (1+i\omega
   \tau_m)}\cdot\frac{R}{d},
\label{GeneralDiffFT}
\end{equation}}
with {$Z(f) = \frac{\kappa}{f^{\gamma/2}-a_i}$} and $\gamma
\geq 1$.  Here, {$d$ is the
  distance between the recording site and the source (approximately
  1~mm in these experiments}).

In the particular case of a Warburg impedance ($\gamma =1$), we 
have:
{{
\begin{equation}
 F_T^{(diff)}(f) = \frac{R_m~f}{\hat{\kappa}_{w}(1+i\omega
  \tau_m)}\cdot\frac{R}{d},
\label{WarburgDiffFT}
\end{equation}}
where {$\hat{\kappa}_{w}$} is a complex constant {because a Warburg
  impedance is such that $Z(f) = \frac{a+ib}{\sqrt{f}}$. 
  For a quasi-resistive medium $Z(f) = \frac{\kappa}{f^{\gamma}-a_i}$} with 
$\gamma$ very close to zero, we have:
{
\begin{equation}
 F_T^{(diff)}(f) = \frac{R_m}{\hat{\kappa_{r}}\ (1+i\omega
   \tau_m)}\cdot\frac{R}{d},
\label{ResistiveDiffFT}
\end{equation}}
where $\hat{\kappa}_{r}$ is a real constant.}

% Note : Il est certain que le cas parfaitement r\'esistif n'a
% pas de sens physique car il y a n\'ecessairement les effets
% capacitifs entre les cellules. Ainsi, si la d\'ependance en
% fr\'equence est tr\`es petite nous sommes devant le cas
% quasi-resistif. Ce commentaire est important pour pouvoir appliquer
% l'\'equation (22)

% Note : Les param\`etres $a_i$ et  $k_i$ n'ont pas
% n\'ecessairement les m\^emes valeurs num\'eriques dans les cas
% Warburg et quasi-r\'esistif.

{Finally, in the case of a purely capacitive medium,
$\gamma =2$ and we have:
{
\begin{equation}
 F_T^{(diff)}(f) = \frac{R_m~f^2}{\hat{\kappa}_c\ (1+i\omega
   \tau_m)}\cdot\frac{R}{d},
\label{CapacitiveDiffFT}
\end{equation}}
where $\hat{\kappa}_{r}$ is a purely imaginary constant in this
case.}

We now compare Eqs.~\ref{WarburgDiffFT}--\ref{CapacitiveDiffFT} with the
experimental measurements. We have analyzed four neurons in which simultaneous
$V_m$ and (bipolar) $V_{LFP}$ were obtained from rat barrel cortex {\it in
  vivo}.  Because the theoretical estimates are for linear regime activity,
our analysis must avoid any possible interference with spikes, and focus
solely on long periods of subthreshold activity as marked by the grey shaded
boxes superimposed on the $V_m$ and LFP traces (see Fig.~\ref{Inviv1}).  The
bottom panels of Fig.~\ref{Inviv1} show the PSD of the $V_m$ and of $V_{LFP}$,
which display similar frequency-scaling exponents.  {Note that the
  exponent values of V$_m$ activity lie within the range identified for other
  anesthetic conditions, for which the V$_m$ exponent varies between -2 and -3
  \cite{Bous2009,Ru2005}}

To compute the transfer function from these data sets, {we evaluate
the ratio between the absolute value of the Fourier transform of
both $V_m$ and LFP} as shown in Fig.~\ref{TransData}A (light grey
curves), which corresponds to the data of Fig.~\ref{Inviv1}. As
suggested by this similar scaling (bottom panel of
Fig.~\ref{Inviv1}), the transfer function of the data
$F_T^{(diff)}$ has a mean value that is approximately constant
(slope zero) for a large frequency range ({about 10~Hz to 500~Hz})
(Fig.~\ref{TransData}A-B).

{We performed a constrained nonlinear least square fit for the three transfer
  functions (Eqs.~\ref{WarburgDiffFT}, \ref{ResistiveDiffFT} and
  \ref{CapacitiveDiffFT}) to the calculated data transfer function
  $F_T^{(diff)}$ for the frequency range between 3~Hz and 500~Hz.} {The
  membrane time constant was constrained to physiological range, $\tau_m \in
  [5~ms, 50~ms]$, whereas the lumped parameter, $\alpha=\frac{R_m R}{\hat{k}
    d}$, was allowed to vary for a large set, $\alpha \in [0, 10^3]$.  Note
  that we chose not to fit all the parameters ($R_m$, $R$, $\hat{k}$ and $d$)
  as some of these parameters are related (e.g. $R_m$ and $\tau_m$) and there
  is an indeterminacy about these parameters because they are lumped.  In
  addition, some parameters will vary from experiment to experiment. For
  instance, electrode coefficients, size of the measured cell, diffusion
  constants (embedded in the parameter $\hat{k}$, as indicated by Eqs.~53-54
  in~\cite{BedDes2009a}) will vary across different experiments.  To fit all
  these parameters, a different experimental protocol would be required to
  gather the necessary data to disambiguate them. Hence, we decided to lump
  them into the single parameter $\alpha$ and also for the fitting purpose we
  consider $\alpha$ and $\tau_m$ to be independent.}

{To ensure soundness in the parameter fitting, we employed two
different averaging techniques to the transfer function
$F_T^{diff}$ (see Methods).  Polynomial averaging techniques (see
Appendix~B) are known to produce robust results when the variance
of the signal is very high~\cite{PressEtAll}.  This method is
applied here in the frequency domain, and virtually suppresses all
of the variance of the transfer function (Fig.~\ref{TransData}A-B,
main plots; see the dashed magenta curves).  In addition, we also
used a moving average window procedure (see Methods), which
markedly reduces the variance (Fig.~\ref{TransData}, insets; see
dark grey shaded curves superimposed on the original PSD in light
grey).}

{{The theoretical expressions for the transfer function were fit to
the transfer functions reconstructed from the experiments.}  These
fits were performed for a Warburg-type medium
(Fig.~\ref{TransData}, solid lines), for a resistive medium (dashed
lines) and for purely capacitive medium (dashed red lines). 
Constrained to the frequency range of 3~Hz to 500~Hz, the fit was
always markedly better for Warburg type impedances, for both cells
shown in Fig.~\ref{TransData} and for both methods, as shown by the
values of the residuals (see Table~1).  Note that for a
resistive-type medium, as well as for a purely capacitive medium,
the parameter estimation {(in particular for $\tau_m$)} always
reached the boundaries of the constrained values, which is
indicative that the obtained minimum in parameter space was far
from optimal {(even if the time constants may coincidentally
reach realistic values)}. To confirm this, we allowed an
unconstrained parameter fit and we observed that although the
resistive-type would seemingly improve the fit (i.e., the initial
plateau of the resistive-type curve would shift towards the
10-500~Hz range), it would however provide unrealistic parameter
ranges.  For the purely capacitive case, it was always impossible
to achieve a good fit (i.e., slope zero).  This is to be expected
as the numerator of Eq.~\ref{CapacitiveDiffFT} grows quadratically
and faster than the denominator, hence it behaves like the identity
function, {$y = f$}, as frequency tends to infinity.  Also note
that the polynomial averaging algorithm of Appendix~B always gave
better results (see values of the residuals in Table~1), although
performed on a dataset {of} higher variance (compare the grey lines
in the main plots and insets of Fig.~\ref{TransData}).}

{Note that the large difference in the residuals between the
polynomial and moving average fits, in particular for the wargburg
impedance, can be explained by the fact that the moving average
does not fully eliminate the variance of the signal, while all the
variance is virtually removed by the polynomial method (see magenta
curves in Fig.~\ref{TransData}A-B).  The error increases even
further for larger frequencies as the slope starts to change,
possibly due to other phenomena, which the model does not capture. 
Also note that polynomial and moving average methods were compared
in a previous study (see chapter 14 and Fig.~14.8.1 in
\cite{PressEtAll}), which should be consulted for details. We stress
that for the Warburg type medium that values obtained for $\tau_m$
and $\alpha$ are consistent with the data and theoretical
predictions. In particular, the theory with Warburg impedance
predicts that the values of $\alpha$ should be negligible for
frequencies greater than $1Hz$ (see Eqs.~53-54 in Appendix of
ref.~\cite{BedDes2009a}).}

{Similar results were also obtained for two other cells of the same database
  (not shown).  The same results were also found by performing a similar
  analysis on another data set consisting of simultaneous intracellular and
  LFP recordings in awake cats (not shown; data from \cite{Steriade2001a},
  courtesy of Igor Timofeev, Laval University, Canada).}

%-------------------------------------------------------------------------------------------------------------------------------------
%    Figure
%-------------------------------------------------------------------------------------------------------------------------------------

\begin{figure}[h!] \centering

\includegraphics[width=\columnwidth]{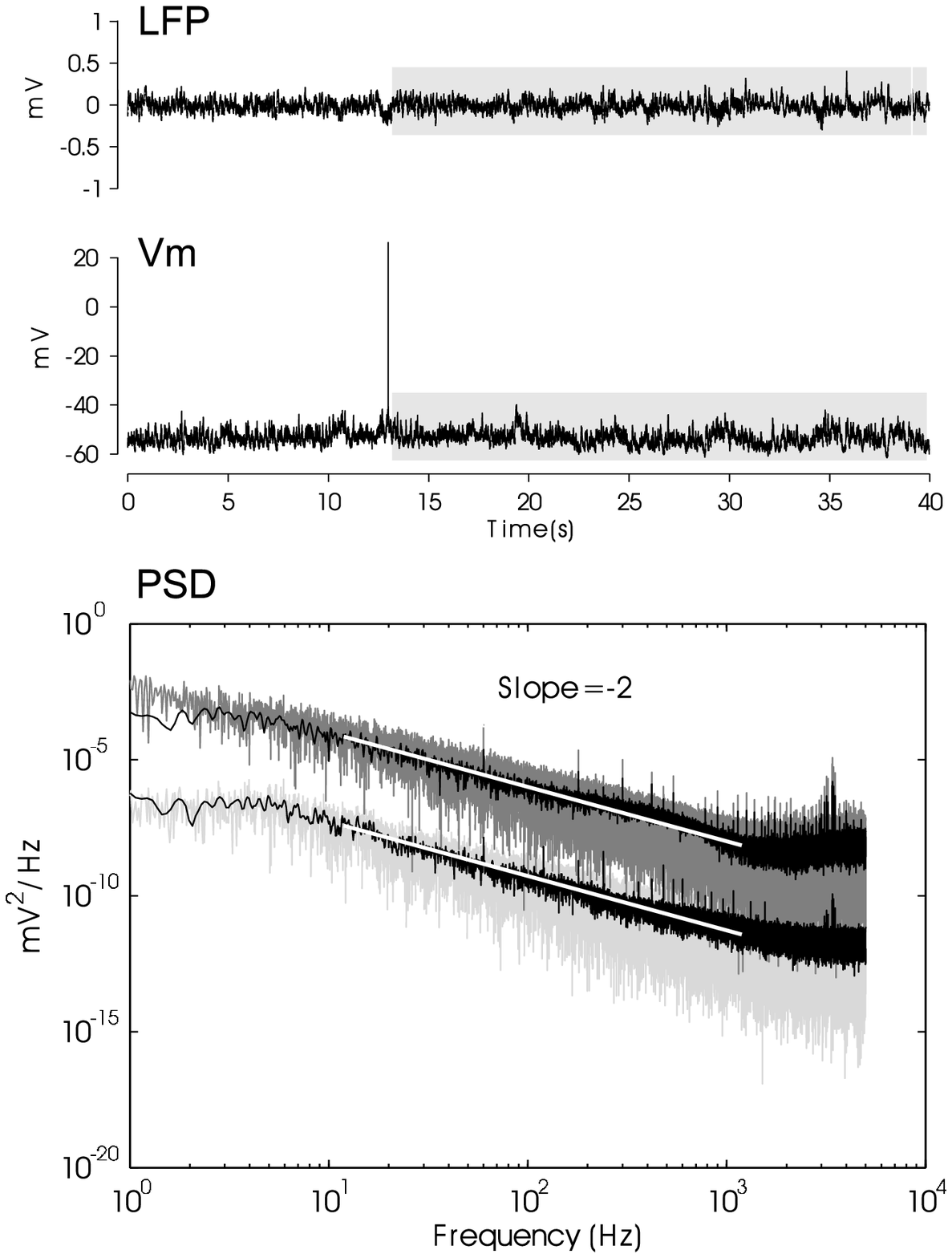}

\caption{Power spectra of simultaneous intracellular and LFP
recordings in desynchronized states {\it in vivo}.  The top panels
depict the time series of simultaneously recorded bipolar LFP and
$V_m$ from rat barrel cortex in light anesthesia with low-amplitude
desynchronized EEG (all recordings at zero current). The shaded
grey box indicate the time period {($T\approx 38.69 ~s$)} of
subthreshold activity selected for analysis.  The bottom panels are
the calculated Power spectra (PSD) of both V$_m$ (top plot in dark
grey) and LFP (bottom plot in light grey), which show similar
scaling.  Superimposed, in black, are the moving average PSD with a
window of $\sim 7.7sec$ (see Methods).  This procedure results in a
PSD with reduced variance but also reduced frequency resolution. 
The total number of points analyzed was N=386900 (between $2^{18}$
and $2^{19}$). {The similar scaling between V$_m$ and LFP is
highlighted by the overlaid white lines that have exactly
$slope=-2$.  In particular, the V$_m$ scales with an exponent
comprised between -2 and -2.4, while the LFP exponents range
between -1.9 and -2.6.}}

\label{Inviv1}
\end{figure}
%-------------------------------------------------------------------------------------------------------------------------------------

%-------------------------------------------------------------------------------------------------------------------------------------
%    Figure
%-------------------------------------------------------------------------------------------------------------------------------------

\begin{figure}[h!]
  \centering
\includegraphics[width=\columnwidth]{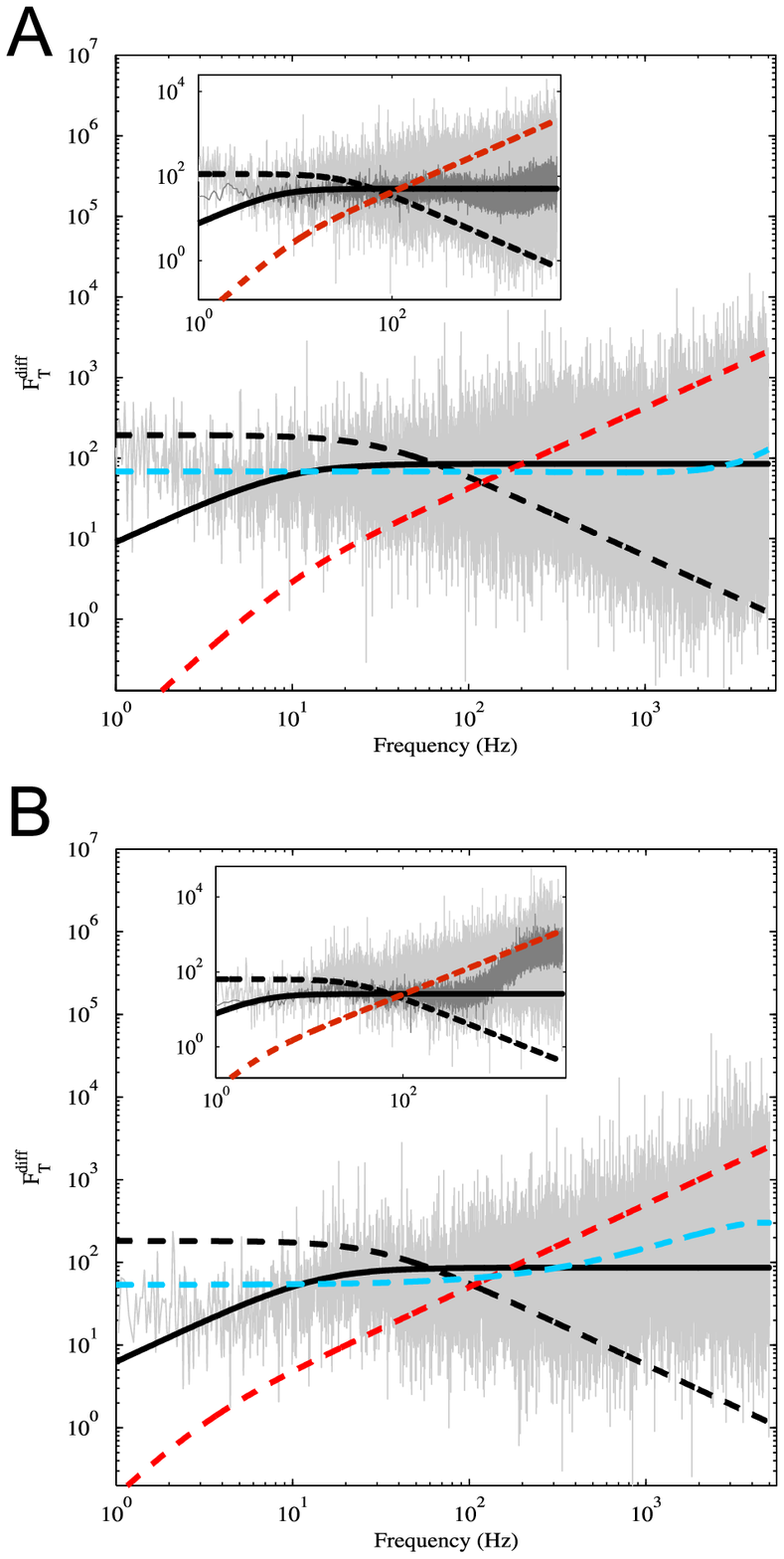}

\caption{{Transfer function $F_T^{(diff)}$ computed from
experimental data.  The top panel corresponds to the cell shown in
Fig.~\ref{Inviv1} and the bottom panel to a different cell.  The
experimentally calculated $F_T^{(diff)}$ (shown in light grey)
shows an average slope of zero for frequencies between about {3~Hz
and 500~Hz}, and is compared to the best fits using a Warburg-type
medium (solid black line), a resistive medium ({black} dashed
line) and purely capacitive medium (red dashed line).  Two
different methods were used to calculate the best nonlinear least
square fit, a polynomial averaging algorithm (main plots,
{third-order polynomial average shown as dashed blue curves};
see Appendix~B for details) and a moving average method which
reduces the variance of the PSD (insets; see Methods). In both
cases, the fit was constrained to frequencies between 3 and 500~Hz.
The parameters of the respective fits are given in Table~1.}}

\label{TransData} 
\end{figure}

%-------------------------------------------------------------------------------------------------------------------------------------

%-------------------------------------------------------------------------------------------------------------------------------------
%    Table
%-------------------------------------------------------------------------------------------------------------------------------------

\begin{table}[h!]
\centering

{
\begin{tabular}{l|l|l|l|l}
Cell    & Impedance type & $\epsilon$       & $\tau_m$ & $\alpha$\\
 & & & \\
\hline
 & & & \\
Cell 1  & Warburg        & $3.4\times 10^2$ & 17.5~ms & 1.43\\
        & Resistive      & $5\times 10^3$   & 5~ms  & 194.94 \\
        & Capacitive     & $1\times 10^8$   & 15~ms  & 0.001 \\
 & & & \\
Cell 1* & Warburg        & $3.2\times 10^5$ & 14.8~ms & 1.22 \\
        & Resistive      & $1.3\times 10^6$ & 5~ms & 111.74 \\
        & Capacitive     & $4.3\times 10^7$ & 15~ms & 0.010\\
 & & & \\
Cell 2  & Warburg        & $3.4\times 10^2$ & 24.5~ms & 1.3\\
        & Resistive      & $5\times 10^3$   & 5~ms & 186.10\\
        & Capacitive     & $7.8\times 10^7$ & 50~ms & 0.004 \\
 & & & \\
Cell 2* & Warburg        & $1.4\times 10^6$ & 24.4~ms & 1.28\\
        & Resistive      & $1.7\times 10^6$ & 5~ms & 63.84\\
        & Capacitive     & $9.3\times 10^6$ & 50~ms & 0.002\\
\end{tabular}
}

\caption{{Parameters for the fitting of the transfer function to
    experimental data for different types of extracellular impedance.
    { $\epsilon = ||y - \hat{y}(\tau_m, \alpha) ||_2$: squared
      2-norm of the fit residual, where $y$ denotes the data and
      $\hat{y}(\tau_m, \alpha)$ represents the various models (Warburg,
      Resistive and Capacitive).  $\tau_m$: membrane time constant and
      $\alpha = \frac{R_m R}{\hat{k}d}$ (refer back to equations in the
      main text), which are obtained by the fitting procedure}. Cell~1
    and Cell~2 correspond to a) and b) in Fig.~\ref{TransData}; Cell~1*
    and 2* refer to the values obtained in the same cells using the
    moving average method (Fig.~\ref{TransData}, insets).} } 

\end{table}

%-------------------------------------------------------------------------------------------------------------------------------------

% ------------------------------------------------------------
%  Discussion
% ------------------------------------------------------------
%\clearpage
\section{Discussion}

In this paper, we have examined the transfer function between intracellular
and extracellular potentials.  {By using a mean-field approximation {in
    Fourier frequency space}, we derived a method allowing us to obtain an
  expression relating the LFP with the intracellular V$_m$ activity.}  The
main theoretical finding is that this transfer function ({which does not
  depend on the frequency spectrum of current sources)} takes very different
forms according to the type of frequency dependence of the extracellular
medium, and thus could be used as a means to estimate which type of frequency
dependence (if any) is most consistent with experiments.  Second, we have
{applied this formalism to intracellular recordings in desynchronized EEG
  states, for which the mean-field approximation should best apply.  We found
  that,} in rat barrel cortex, the extracellular medium seems frequency
dependent with a Warburg type impedance.

{One key assumption of the present formalism is that individual
synaptic currents sources are uncorrelated.  There is ample
evidence that this is the case for EEG-desynchronized states, as
shown by the low levels of correlation between simultaneously
recorded
units~\cite{Contreras96,Des1999,Gawne93,Steriade2001b,Zohary94}, or
by the low correlations between multi-site LFPs~\cite{Des1999}. 
However, there are limits to this assumption.  First, dual
recordings in awake mice barrel cortex showed that the subthreshold
activities of neurons can display periods of significant
correlation, even with desynchronized EEG~\cite{Poulet2008}. 
Indeed, a certain level of correlation is unavoidable from the
redundant connectivity of neurons in cortex, although the activity
itself can produce negative correlations which may cancel the
effect of redundant connectivity~\cite{Renart2010}.  Second, it is
evident that the activity cannot be totally decorrelated, otherwise
the EEG amplitude would be close to zero~\cite{Gold2006}. 
Nevertheless, the amplitude of the EEG during desynchronized states
is considerably lower than during synchronized activity (such as
slow waves), which is associated to a general decorrelation of
neurons~\cite{Contreras96}, and the present formalism should be
applicable to such desynchronized states.}  {Further studies
should consider these points to build more realistic mean-field
models of desynchronized states, which would lead to more realistic
transfer functions.}

In a previous investigation \cite{BedDes2009a}, we have shown theoretically
that several physical phenomena can lead to frequency dependence of the
extracellular medium: ionic diffusion and membrane polarization.  The former
predicts an impedance of Warburg type ({$Z \sim 1 / \sqrt{f}$}), while
the latter predicts a capacitive-type impedance {($Z \sim 1/f$)}
\cite{Bed2006b}.  These two phenomena can also explain different experimental
observations: the frequency dependent conductivity observed experimentally in
brain tissue \cite{Gabriel1996} can be reproduced by a combination of these
two mechanisms.  Recent measurements from monkey cortex suggesting resistive
medium \cite{Logo2007} can be explained by the fact that the influence of
diffusion was avoided in that case. {This technique is based on the
  saturation effect (Geddes effect, which is represented by Zener diodes in
  \cite{Logo2007}), which greatly diminishes the concentration gradient around
  the electrode, such that the ionic diffusion is more limited.}  A Warburg
type impedance was also found to account for the $1/f$ power spectral
structure of LFPs (see details in \cite{BedDes2009a}).

The present results are consistent with this analysis.  The transfer functions
measured here for 4 cells are all consistent with the Warburg type impedance
of ionic diffusion {up to 500~Hz.  The other type of extracellular impedances
  mentioned above, either purely resistive or purely capacitive, could not fit
  the data (see Fig.~\ref{TransData}).  While these results seem to rule-out
  purely resistive or capacitive media, there is still a possibility that they
  apply outside the 3-500~Hz frequency range.  For example, polarization
  phenomena, which can be modeled as a capacitive effect with a low cutoff
  frequency \cite{Bed2006b}, may contribute to the low frequency range (below
  10~Hz).  Further theoretical and experimental work is needed to investigate
  these aspects.  In particular, experiments should be carried with controlled
  current sources as close as possible to the biological current sources, for
  example using micropipettes.}

{It is important to keep in mind that the present method derives
from a mean-field approach {in frequency space}, and thus relies on
the assumption that individual current sources are independent. 
This justifies the use of desynchronized EEG states, in which
synaptic current sources are expected to have very low correlation.
{The simulations presented in Fig.~\ref{desynch} {(bottom
panels)} show that the LFP predicted by a compartmental dendritic
model is virtually identical to that obtained by individual
compartments, which suggests that considering a set of independent
sources is not a bad approximation.}  However, if strong
correlations occur, such as during synchronized population
activities, the current sources may no longer be considered as
independent, and another formalism should be used.}

A recent study on modeling extracellular action potentials
\cite{PetEin2008} showed that the cable structure of neurons can
also cause low-pass filtering, because the return current is itself
filtered by the membrane capacitance.  This is a clear example
where the correlation between current sources cannot be neglected.
{However, we showed here that such a contribution is negligible for
{\it in vivo} conditions (Fig.~\ref{desynch}A,B), presumably
because the axial currents are very small compared to the intense
synaptic currents.} So far, the only plausible physical cause to
explain the observed $1/f$ filtering {under {\it in vivo}
conditions} is ionic diffusion {(for frequencies up to 500~Hz)}.
{It is possible that for states of reduced synaptic activity, the
filtering due to morphology plays a role, although this still needs
to be demonstrated experimentally } Further studies should
investigate these aspects by constraining these different theories
by appropriate experiments.}

{One major criticism to the previous measurement techniques
  \cite{Ranck63,Gabriel1996,Logo2007} is that they use current intensities of
  one or several order of magnitude larger than biological sources, and these
  currents evidently interact with the medium very differently as natural
  sources.  The present method has the advantage of not suffering from this
  limitation, because it is using only passive recordings of physiological
  signals, with no need of injecting currents.  This method should therefore
  be considered as complementary to direct measurements.}

Finally, the expression given by Eqs.~\ref{GeneralMonoFT} and
\ref{GeneralDiffFT} could be used to directly estimate the impedance of the
extracellular medium as a function of frequency {(still within the mean-field
  approximation)}.  We did not attempt this type of approach here, but instead
considered different hypotheses concerning the impedance of the medium.  {The
  present analysis reported here for 4 cells was also confirmed by using two
  cells from another database of intracellular recordings in desynchronized
  EEG states in awake animals (courtesy of I.~Timofeev, Laval University,
  Canada), which also indicated a Warburg type impedance (not shown).  The
  same approach should be extended to a much larger sample of cells and brain
  states, to provide a full estimate of the impedance spectrum of the medium.}
The present results therefore must be considered as preliminary and must be
confirmed by using further analyses of simultaneously recorded LFPs and
intracellular recordings {during desynchronized EEG states} {\it in vivo}.

% ------------------------------------------------------------
%  Appendix
% ------------------------------------------------------------
%\clearpage

\section*{Appendix A: impedance for non-ideal membranes}

In this section, we derive the expressions for the impedance of
non-ideal membranes, which take into account that the membrane
capacitance cannot be charged instantaneously (see Bedard and
Destexhe, 2008).  Still within the linear regime and for a spherical
source, we have:
{
\begin{eqnarray}
 I_r &=& \sum_{i=1}^{N}g_i~(t,V_m)( V_m(t)-E_i)\nonumber  \\ \nonumber \\
 I_c &=& C_{m}\frac{d V_c}{dt}\nonumber \\ \\
I_m &=& I_r +I_c \nonumber \\\nonumber  \\
V_m &=& V_c + R_{\mbox{\begin{tiny}MW\end{tiny}}}C_m\frac{d V_c}{dt}=V_c + \tau_{\mbox{\begin{tiny}MW\end{tiny}}}\frac{d V_c}{dt}\nonumber 
\end{eqnarray}}
where all parameters have the same definition as in the main text,
except for {$r_{\mbox{\begin{tiny}MW\end{tiny}}}$}, which is the 
Maxwell-Wagner resistance which gives the non-ideal aspect of the membrane 
capacitance.  The associated time constant, 
{$\tau_{\mbox{\begin{tiny}MW\end{tiny}}}$}, is also known as
``Maxwell-Wagner time''.

In the linear regime, we have
{
\begin{eqnarray}
 \Delta I_r &=& \sum_{i=1}^{N}g_i~(\avg{V_m}|_t)
  \Delta V_m\nonumber  \\ \nonumber \\
 \Delta I_c &=& C_{m}\frac{d \Delta V_c}{dt}\nonumber \\ \\
 \Delta I_m &=& \Delta I_r +\Delta I_c \nonumber \\\nonumber  \\
 \Delta V_m &=& \Delta V_c + \tau_{\mbox{\begin{tiny}MW\end{tiny}}}
\frac{d \Delta V_c}{dt}\nonumber 
\end{eqnarray}}
Thus, in these conditions, the system of equations associated to the
membrane is linear with time-independent coefficients.

By expressing the variation of current produced by the cell as a
function of the variation of membrane voltage, in Fourier space, we
obtain:
{
\begin{eqnarray}
 \Delta I_r(f) &=& G_m  \Delta V_m(f)\nonumber  \\ \nonumber \\
 \Delta I_c(f) &=& i\omega C_{m} \Delta V_c(f)\nonumber \\ \\
 \Delta I_m(f) &=& \Delta I_r(f) +\Delta I_c(f) \nonumber \\\nonumber  \\
 \Delta V_m(f) &=& \Delta V_c(f) + 
i\omega \tau_{\mbox{\begin{tiny}MW\end{tiny}}}\Delta V_c(f)\nonumber
\end{eqnarray}}
where $$G_m=\sum_{i=1}^{N}g_{i}$$.

It follows that the membrane impedance is given by:
{
\begin{equation}
 Z_m(f)=\frac{\Delta V_m(f)}{\Delta I(f)} = 
 \frac{R_m}{1+i\frac{\omega \tau_m}{1+i\omega \tau_{\mbox{\begin{tiny}MW\end{tiny}}}}}
\label{ZmN}
\end{equation}}
where $R_m =\frac{1}{G_m}$.  Note that if we set 
{$\tau_{\mbox{\begin{tiny}MW\end{tiny}}}=0$}, we
recover the same expressions for the impedance of ideal membranes, as
considered in the main text.

{\section*{Appendix B: Polynomial averaging algorithm for
frequency-dependent signals}}

{The polynomial averaging technique consists of fitting a
polynomial to the cumulative distribution of the amplitude of the
signal in frequency space.  According to this procedure, one
evaluates} the difference between the data and model via a
minimization problem (in the frequency space) as follows
\begin{equation}
argmin_{\tau_m, \alpha} ||y(f) - \hat{y}(f, \tau_m, \alpha)||, \quad f\in[3Hz, 500Hz],
\end{equation}
where $y(f)$ represents the transfer function (after a Fourier transform or
PSD has been applied). $\hat{y}(\cdot)$ represent the various model transfer
functions (i.e. Warburg, resistive or capacitive) parameterized by $\tau_m$
and $\alpha$. Since the theory we develop only explains changes in the slope
of the mean of $y(f)$, then to a first order approximation the above equation
can be re-written in the following way
\begin{equation}
argmin_{\tau_m, \alpha} ||<y>(f) - \hat{y}(f, \tau_m, \alpha)||, \quad f\in[3Hz, 500Hz]
\end{equation}
where the operator $<\cdot>$ is the mean of the data in frequency domain.  Any
technique can be taken to evaluate the mean such as the moving
average. However, the moving average {does not} completely remove the variance and
is not general enough. Since Fourier transform of the transfer function or its
PSD show a large variance we remove entirely this variance by employing the
following polynomial algorithmic filter.  }

{1. The Fourier transform of the signals is integrated relative to 
frequency:
\begin{equation}
 G(f) =\int_{f_{min}}^{f}F(f')df'
\end{equation}
where $f_{min}$ is the minimal frequency considered with $f \leq
F_{max}$, and $F(f')$ is the signal for which the mean function must
be obtained.  This integration gives a function of frequency which is
very close to the integral of the mean function, which is true for
the signals considered here.} 

2. To smooth the function $G(f)$, a minimum variance fit is performed 
using a third-degree polynomial:
{
\begin{equation}
 G^*(f)=A_3f^3 + A_2f^2+A_1f+A_0 + \mbox{O}(f^4) ~ .
\end{equation}}
{Note that higher-order polynomials can be used to improve 
accuracy. However, in our case we did not observe any gain for
orders larger than three.}

{3. This polynomial was formally derived to find the expression of the mean
function $<F>(f)$
\begin{equation}
 <F>(f)=\frac{dG^*}{df} =3A_3f^2 + 2A_2f+A_1 ~ .
\end{equation}}

{Note that this algorithm is general and does not depend on any
hypothesis concerning the stationarity of the signal because the
average function is calculated in Fourier space.} \\

% ------------------------------------------------------------
%  Acknowledgments
% ------------------------------------------------------------

\section*{Acknowledgments}

Research supported by the Centre National de la Recherche
Scientifique (CNRS, France), Agence Nationale de la Recherche (ANR,
France) and the Future and Emerging Technologies program (FET,
European Union; FACETS project).  Additional information is available
at \url{http://cns.iaf.cnrs-gif.fr}

% ------------------------------------------------------------
%  References
% ------------------------------------------------------------

% Reference list entries should be alphabetized by the last names of
% the first author of each work.
% 
% Journal article
% Harris, M., Karper, E., Stacks, G., Hoffman, D., DeNiro, R., Cruz,
% P., et al. (2001). Writing labs and the Hollywood connection. 
% Journal of Film Writing, 44(3), 213–245.
%
% Book
% Calfee, R. C., & Valencia, R. R. (1991). APA guide to preparing
% manuscripts for journal publication. Washington, DC: American
% Psychological Association.
%
% Book chapter
% O’Neil, J. M., & Egan, J. (1992). Men’s and women’s gender role
% journeys: Metaphor for healing, transition, and transformation. In B.
% R. Wainrib (Ed.), Gender issues across the life cycle (pp. 107–123).
% New York: Springer.
%
% Online document
% Abou-Allaban, Y., Dell, M. L., Greenberg, W., Lomax, J., Peteet,
% J., Torres, M., Cowell, V. (2006). Religious/spiritual commitments
% and psychiatric practice. Resource document. American Psychiatric
% Association.
% http://www.psych.org/edu/other_res/lib_archives/archives/200604.pdf.
% Accessed 25 June 2007.

%\clearpage
%\section*{References}

%\begin{description}

\end{document}